# Strategic Interactions in Academic Dishonesty: A Game-Theoretic Analysis of the Exam Script Swapping Mechanism

Venkat Ram Reddy Ganuthula
*Indian Institute of Technology Jodhpur*
ram@iitj.ac.in

Manish Kumar Singh
*Indian Institute of Technology Roorkee*
mks@hs.iitr.ac.in

October 17, 2025

**Abstract**

This paper presents a novel game-theoretic framework for analyzing academic dishonesty through the lens of a unique deterrent mechanism: forced exam script swapping between students caught copying. We model the strategic interactions between students as a non-cooperative game with asymmetric information and examine three base scenarios: asymmetric preparation levels, mutual non-preparation, and coordinated partial preparation. Our analysis reveals that the script-swapping punishment creates a stronger deterrent effect than traditional penalties by introducing strategic interdependence in outcomes. The Nash equilibrium analysis demonstrates that mutual preparation emerges as the dominant strategy. The framework provides insights for institutional policy design, suggesting that unconventional punishment mechanisms that create mutual vulnerability can be more effective than traditional individual penalties. Future empirical validation and behavioral experiments are proposed to test the model's predictions, including explorations of tapering off effects in punishment severity over time.

# 1 Introduction

Academic dishonesty remains a persistent challenge in educational institutions worldwide, with studies indicating that 60–70% of students engage in some form of cheating during their academic careers. Traditional approaches to deterring academic dishonesty have relied primarily on detection and punishment mechanisms that treat cheating as an individual decision problem, following the economic framework established by Becker (1968). However, the social nature of many forms of academic dishonesty, particularly collaborative cheating during examinations, suggests that a game-theoretic framework may provide deeper insights into the strategic considerations underlying students' decisions to cheat.

This paper introduces and analyzes a novel deterrent mechanism: the forced swapping of exam scripts between students caught copying, requiring each to complete the other's examination. This mechanism, while unconventional, has been discussed in various forms across educational institutions in South Asia and presents unique strategic considerations that differ fundamentally from traditional punishment approaches. Unlike conventional penalties that affect only the individual cheater, script swapping creates a situation of mutual vulnerability where each student's outcome depends critically on the other's preparation level—a form of strategic interdependence that aligns with principles of responsive regulation.

The game-theoretic analysis of this mechanism reveals several important insights. First, the interdependence created by potential script swapping transforms what would otherwise be a dominant strategy (copying from a prepared student) into a risky proposition that depends on detection probability and the expected preparation level of one's partner. Second, the mechanism creates natural incentives for honest preparation that are robust to various student characteristics and institutional contexts. Third, the framework suggests that punishment mechanisms that create strategic complementarities between potential violators may be more effective deterrents than traditional individual-focused penalties.

## 1.1 Contributions

Our contribution to the literature is threefold. First, we provide the first formal game-theoretic analysis of the script-swapping mechanism, establishing conditions under which it serves as an effective deterrent. Building on the game-theoretic foundations established by recent research, we extend the basic model to incorporate realistic features such as heterogeneous student abilities, incomplete information about preparation levels, and repeated interactions across multiple examinations.

Second, we contribute to the broader literature on punishment mechanism design by demonstrating how creating strategic interdependence can achieve deterrence more efficiently than independent penalties of equivalent severity. This extends Becker (1968)'s seminal framework by showing that the structure of punishment, not just its severity, can be a powerful policy instrument.

Third, we derive testable predictions about student behavior under different institutional arrangements and punishment mechanisms, providing a foundation for future empirical work. The remainder of this paper proceeds as follows: Section 2 reviews the relevant literature, Section 3 presents the model setup, Section 4 analyzes base scenarios, Section 5 examines equilibrium properties, Section 6 discusses the findings, Section 7 ex-plores implications, Section 8 outlines future research directions, and Section 9 concludes.

## 2 Literature Review

### 2.1 Economic Analysis of Academic Dishonesty

The economic analysis of academic dishonesty began with Becker (1968)'s seminal work on the economics of crime, which modeled criminal behavior as a rational choice based on expected costs and benefits. Applying this framework to academic settings, Bunn et al. (1992) demonstrated that students' decisions to cheat could be understood as rational responses to incentives, including the probability of detection, severity of punishment, and potential benefits from higher grades. Kerkvliet and Sigmund (1999) extended this analysis by incorporating student characteristics and institutional factors, finding that deterrent effects were strongest when both detection probability and punishment certainty were high.

The limitations of individual decision models became apparent as researchers recognized the inherently social nature of many forms of academic dishonesty. Carrell et al. (2008) documented significant peer effects in cheating behavior, showing that students were more likely to cheat when their peers did so. This finding suggested that game-theoretic models accounting for strategic interactions might provide better explanations than individual optimization models alone. Pascual-Ezama et al. (2015) provided experimental evidence that peer presence constrains dishonest behavior, supporting the notion that social observation creates deterrent effects—a mechanism central to our script-swapping analysis where mutual vulnerability creates strategic incentives for honest behavior.

### 2.2 Game-Theoretic Applications in Education

Game theory has been increasingly applied to educational contexts to understand strategic interactions between students, between students and instructors, and within institutional settings. Lester and Diekhoff (2002) were among the first to explicitly model cheating as a game between students, showing that multiple equilibria could exist depending on social norms and detection technologies. Their pioneering work established the game-theoretic framework for analyzing academic dishonesty, demonstrating that cheating could become endemic when social sanctions were weak.

More recently, Griebeler (2019) developed a three-player game model analyzing strategic relationships between students and professors, examining peer effects on student cheating and emphasizing social information and sorting mechanisms. Young (2020) applied prisoner's dilemma concepts to academic dishonesty, supporting the Nash equilibrium solution concept and providing validation for findings about mutual preparation as a dominant strategy under appropriate parameter conditions. Rigby et al. (2015) developed a game-theoretic model of collaborative cheating that incorporated network effects, showing that the structure of social networks within classrooms significantly influenced equilibrium cheating levels.

## 2.3 Deterrence Theory and Punishment Mechanism Design

The literature on optimal punishment design has evolved from simple cost-benefit analyses to sophisticated mechanism design problems. In educational contexts, traditional approaches have focused on increasing detection probability through proctoring, plagiarism detection software, and honor codes. Kleiman (2009) demonstrated how strategic targeting and interdependence enhance punishment effectiveness more than increased severity alone—a principle directly applicable to script swapping. His work shows that making outcomes interdependent between co-offenders can create stronger deterrent effects than harsher individual punishments.

Recent work has explored alternative deterrent mechanisms beyond traditional penalties. Dee and Jacob (2012) examined the effectiveness of honor codes and found that their impact depended critically on implementation details and institutional culture. Cohn and Maréchal (2018) studied commitment devices that allowed students to voluntarily restrict their own cheating opportunities, finding that such mechanisms could be surprisingly effective. Farjam (2015) studied various punishment mechanisms through simulation, finding that interdependent punishment systems often outperform individual-focused approaches—directly supporting our script-swapping mechanism's strategic complementarity argument.

The concept of "responsive regulation" introduced by Braithwaite (2002) has influenced recent thinking about academic integrity policies. This approach emphasizes graduated responses that begin with education and support before escalating to punitive measures, potentially incorporating tapering off in punishment severity for repeat offenders who show signs of reform. However, little work has examined punishment mechanisms that create strategic interdependence between violators, as in the script-swapping scenario we analyze—representing a significant gap in the literature that our paper addresses.

## 2.4 Behavioral Considerations and Student Heterogeneity

While our model assumes rational, forward-looking agents, behavioral economics research has documented important deviations from perfect rationality in dishonesty contexts. Ariely (2012) shows that people exhibit systematic behavioral patterns when engaging in dishonest behavior, including self-deception and moral disengagement that may not align with pure cost-benefit calculations. Wang and Zhang (2022) demonstrate how individual differences in risk preferences and moral reasoning affect cheating behavior, suggesting that student heterogeneity extends beyond ability differences to include psychological factors.

These behavioral insights suggest that the deterrent effect of script swapping may be strengthened by psychological factors beyond our formal model. The prospect of being forced to complete another student's work may trigger loss aversion, guilt, or shame—emotions that could amplify the mechanism's effectiveness.

## 2.5 Empirical Evidence and Cultural Context

Empirical studies have documented widespread academic dishonesty across institutions and countries. McCabe et al. (2012) surveyed over 70,000 students across 150 institutions, finding that 68% admitted to some form of cheating. The prevalence varies by discipline, with business and engineering students reporting higher rates than those in humanities.

Cultural factors also play a significant role. Magnus et al. (2002) compared cheating rates across four countries, finding substantial variation that correlated with cultural attitudes toward cooperation and competition. Bernardi et al. (2017) extended this analysis to multiple countries, documenting how institutional and cultural contexts shape both the prevalence and forms of academic dishonesty. This cross-cultural variation is particularly relevant for script swapping, which has been discussed primarily in South Asian educational contexts, suggesting that cultural acceptance may be an important implementation consideration.

## 2.6 Positioning of This Study

Our work contributes to this literature by providing the first formal analysis of an interdependent punishment mechanism in academic contexts. While previous research has examined peer effects, game-theoretic interactions, and punishment design, none has analyzed how creating mutual vulnerability through script swapping affects strategic incentives. Our model bridges rational choice theory with game-theoretic strategic interaction, providing testable predictions that can be evaluated through experimental methods or field studies when institutions implement such policies.

# 3 Model Setup and Assumptions

We model the examination scenario as a two-player non-cooperative game. Let players be denoted as Student $A$ and Student $B$, indexed by $i \in \{A, B\}$. Each student must make two sequential decisions: first, whether to prepare for the examination (study) or not prepare (slack), and second, whether to attempt copying during the examination.

## 3.1 Strategy Space and Player Characteristics

The preparation decision is made simultaneously by both students before the examination, represented by strategy $s_i \in \{P, NP\}$, where $P$ denotes "Prepare" and $NP$ denotes "Not Prepare." The copying decision is conditional on the preparation level and beliefs about the other student's preparation. For tractability, we initially assume that students who have not prepared will always attempt to copy if they believe the other student has prepared, while prepared students do not copy.

Let $\theta_i \in [0, 1]$ represent student $i$'s inherent ability level, where higher values indicate greater academic capability. Following Spence (1973)'s signaling framework, ability affects both the cost and effectiveness of preparation. The preparation effort cost is given by $c_i = c/\theta_i$, where $c > 0$ is a base effort cost parameter. This specification captures the intuition that more able students find preparation less costly.

## 3.2 Payoff Structure

The examination yields a raw score $r_i$ based on the student's preparation level and ability:

$$r_i = \begin{cases} \theta_i \cdot g & \text{if } s_i = P \\ \theta_i \cdot g \cdot \alpha & \text{if } s_i = NP \end{cases} \quad (1)$$

where $g$ represents the maximum possible grade and $\alpha \in (0, 1)$ captures the fraction of the exam that can be completed without preparation based solely on ability.

If a student attempts to copy and is not detected, they receive the maximum of their own score and the copied score. The probability of detection is given by $p \in (0, 1)$, which depends on institutional factors such as proctoring intensity, classroom layout, and examination procedures.

## 3.3 The Script-Swapping Mechanism

When students are caught copying, the script-swapping mechanism is triggered. Each student must complete the remainder of the other student's examination script. We specify the payoff function for student $i$ as:

$$U_i = \begin{cases} r_i - c_i & \text{if no copying occurs} \\ \max(r_i, r_j) - c_i & \text{if copying occurs and not detected} \\ \phi(r_i, r_j, \theta_i, \theta_j) - c_i & \text{if copying detected and scripts swapped} \end{cases} \quad (2)$$

The swap payoff function $\phi(\cdot)$ captures the complex interaction when scripts are exchanged:

$$\phi(r_i, r_j, \theta_i, \theta_j) = \beta \cdot r_j + (1 - \beta) \cdot \theta_i \cdot g \cdot \gamma \quad (3)$$

where $\beta \in (0, 1)$ represents the portion of the exam completed before swapping, and $\gamma \in (0, 1)$ reflects the difficulty of completing someone else's partially finished exam.

## 3.4 Information Structure and Solution Concept

We consider three information scenarios: (i) Complete information where both students know each other's ability levels and preparation strategies; (ii) Incomplete information where students know their own ability but have only probabilistic beliefs about the other's ability and preparation; and (iii) Asymmetric information where one student knows more about the examination format or the other student's characteristics. For the base case analysis, we focus on the complete information scenario.

The game proceeds as follows: (1) Nature determines student abilities ($\theta_A$, $\theta_B$); (2) Students simultaneously choose preparation strategies ($s_A$, $s_B$); (3) Examination begins and students observe each other's apparent preparation; (4) Copying decisions are made; (5) Detection occurs with probability $p$; (6) Payoffs are realized. We solve for subgame perfect Nash equilibrium using backward induction, requiring that strategies constitute mutual best responses at every decision point.

## 3.5 Key Assumptions

For analytical tractability, we maintain the following assumptions: (A1) Students are risk-neutral expected utility maximizers; (A2) The game structure, payoff functions, and detection probability are common knowledge; (A3) Students cannot make binding agreements about preparation or copying; (A4) Students either prepare fully or not at all (no partial preparation in base model); (A5) When cheating is detected, it is discovered immediately during the exam. We relax some of these assumptions in extensions discussed in Section 8.

# 4 Base Case Analysis

We analyze three fundamental scenarios that characterize the strategic landscape of academic dishonesty under the script-swapping mechanism.

## 4.1 Asymmetric Preparation Scenario

Consider the scenario where Student $A$ has prepared ($s_A = P$) while Student $B$ has not ($s_B = NP$). This represents the classic free-rider problem in academic settings. Under traditional punishment, Student $B$'s expected payoff from copying is:

$$EU_B^{copy} = (1-p) \cdot \theta_A \cdot g + p \cdot 0 = (1-p) \cdot \theta_A \cdot g \qquad (4)$$

With the swapping mechanism, the expected payoff becomes:

$$EU_B^{swap} = (1-p) \cdot \theta_A \cdot g + p \cdot [\beta \cdot \theta_A \cdot g + (1-\beta) \cdot \theta_B \cdot g \cdot \gamma] \qquad (5)$$

**Proposition 1** (Deterrent Effect of Script Swapping). *The script-swapping mechanism reduces the expected benefit from copying compared to traditional penalties if and only if:*

$$p \cdot [1 - \beta - (1-\beta) \cdot \gamma] > p_0 \cdot [1 - \tau] \qquad (6)$$

*where $p_0$ is the detection probability under traditional penalties and $\tau$ is the traditional penalty parameter.*

*Proof.* The deterrent effect requires that $EU_B^{swap} < EU_B^{trad}$ where traditional punishment gives $(1-p_0)\theta_A g + p_0 \tau \theta_A g$. Comparing expected payoffs and rearranging yields condition (7). □

The key insight is that the swapping mechanism reduces the expected benefit from copying because $\phi(r_B, r_A, \theta_B, \theta_A) < \theta_A \cdot g$ under reasonable parameters. The unprepared student must complete a partially finished exam in an unfamiliar style, yielding a lower grade than successful copying would provide. This creates a "mutual vulnerability" effect absent in traditional penalties.

For Student $A$, who prepared, being forced to complete Student $B$'s largely empty script is particularly costly:

$$EU_A^{swap} = (1-p) \cdot \theta_A \cdot g + p \cdot [\beta \cdot \theta_B \cdot g \cdot \alpha + (1-\beta) \cdot \theta_A \cdot g \cdot \gamma] - c_A \qquad (7)$$

This creates a strong incentive for prepared students to avoid situations where copying might occur, potentially leading to strategic behaviors such as concealing one's preparation level or choosing seating to minimize copying opportunities.

## 4.2 Mutual Non-Preparation

When neither student has prepared ($s_A = NP$, $s_B = NP$), the situation resembles a coordination failure. Both students would benefit from preparation, but neither has incentive to prepare if they expect the other to slack. Copying provides no benefit since neither student has meaningful answers to copy. The expected payoffs are:

$$EU_A = EU_B = \theta_i \cdot g \cdot \alpha \qquad (8)$$

If scripts are swapped after attempted copying, the payoffs remain essentially unchanged:

$$EU_i^{swap} = \beta \cdot \theta_j \cdot g \cdot \alpha + (1-\beta) \cdot \theta_i \cdot g \cdot \gamma \qquad (9)$$

**Proposition 2** (Mutual Non-Preparation Equilibrium)**.** *Mutual non-preparation (NP, NP) constitutes a Nash equilibrium if and only if:*

$$\theta_i \cdot g \cdot (1 - \alpha) < c_i + p \cdot \theta_i \cdot g \cdot [1 - \beta \cdot \alpha - (1 - \beta) \cdot \gamma] \tag{10}$$

*Proof.* For (*NP, NP*) to be a Nash equilibrium, neither student should have incentive to deviate to *P*. The left side represents the benefit from preparation (increased grade), while the right side represents the cost plus expected loss from potential script swapping if the other student attempts to copy. □

This condition reveals that mutual non-preparation is more likely when: (i) preparation costs are high, (ii) detection probability is high (paradoxically, very high detection can sustain low-preparation equilibria by deterring copying), or (iii) the swapping mechanism is particularly punitive (low $\beta$ or $\gamma$). The interesting dynamic here is that mutual non-preparation represents a Pareto-inefficient equilibrium where both students would be better off if both prepared, but unilateral preparation creates vulnerability to free-riding.

## 4.3 Coordinated Partial Preparation

The most interesting scenario involves coordinated partial preparation, where students divide the curriculum and each prepares different sections. This represents a form of collaborative learning that walks the line between legitimate cooperation and academic dishonesty. Let the exam consist of two equal parts, and assume each student prepares one part thoroughly. Their individual scores without copying would be:

$$r_i^{partial} = \theta_i \cdot g \cdot (0.5 + 0.5 \cdot \alpha) \tag{11}$$

With successful copying, both could achieve:

$$r^{coordinated} = \theta_{max} \cdot g \tag{12}$$

where $\theta_{max} = \max(\theta_A, \theta_B)$.

Under the swapping mechanism, if students have complementary preparation, swapping might actually maintain or even improve outcomes:

$$\phi^{coordinated} = 0.5 \cdot \theta_j \cdot g + 0.5 \cdot \theta_i \cdot g = 0.5 \cdot g \cdot (\theta_i + \theta_j) \tag{13}$$

**Proposition 3** (Coordinated Partial Preparation)**.** *Coordinated partial preparation dominates full individual preparation if:*

$$p \cdot 0.5 \cdot g \cdot (\theta_i + \theta_j) + (1 - p) \cdot \theta_{max} \cdot g > \theta_i \cdot g - c_i \tag{14}$$

*and the cost of partial preparation is sufficiently low.*

This scenario reveals an important limitation of the swapping mechanism: it may not deter all forms of collaborative cheating, especially when students can maintain plausible deniability about their coordination. This finding aligns with research showing that network structure affects collaborative cheating patterns.

# 5 Equilibrium Analysis and Comparative Statics

## 5.1 Payoff Matrix and Equilibrium Characterization

To fully characterize the Nash equilibria of our game, we construct the normal form representation for the preparation stage, incorporating the expected outcomes from the subsequent copying and potential swapping stages. For symmetric students with $\theta_A = \theta_B = \theta$, the payoffs are:

- Both Prepare $(P, P)$: $\pi_i^{PP} = \theta \cdot g - c/\theta$

- Asymmetric $(P, NP)$:

$$\pi_A^{PNP} = (1-p) \cdot \theta \cdot g + p \cdot [\beta \cdot \alpha + (1-\beta) \cdot \gamma] \cdot \theta \cdot g - c/\theta$$
$$\pi_B^{PNP} = (1-p) \cdot \theta \cdot g + p \cdot [\beta + (1-\beta) \cdot \gamma] \cdot \theta \cdot g$$

- Both Don't Prepare $(NP, NP)$: $\pi_i^{NPNP} = \alpha \cdot \theta \cdot g$

**Theorem 1** (Unique Nash Equilibrium). *For symmetric students, the unique Nash equilibrium is mutual preparation $(P, P)$ if and only if:*

$$p > \bar{p} = \frac{c/\theta - \theta \cdot g \cdot (1-\alpha)}{\theta \cdot g \cdot [1 - \beta \cdot \alpha - (1-\beta) \cdot \gamma]} \tag{15}$$

*Proof.* For $(P, P)$ to be a Nash equilibrium, we need $\pi_i^{PP} > \pi_i^{NPP}$. This yields:

$$\theta \cdot g - c/\theta > (1-p) \cdot \theta \cdot g + p \cdot [\beta + (1-\beta) \cdot \gamma] \cdot \theta \cdot g \tag{16}$$

Rearranging gives the threshold $\bar{p}$. Uniqueness follows from verifying that $(NP, NP)$ cannot be an equilibrium when this condition holds. □

This threshold has important policy implications. It shows that the script-swapping mechanism requires a minimum detection probability to ensure honest preparation, but this threshold is notably lower than what traditional penalties would require for the same deterrent effect.

## 5.2 Comparative Statics

We now examine how equilibrium outcomes change with key parameters. Differentiating the threshold with respect to the cost parameter:

$$\frac{\partial \bar{p}}{\partial c} = \frac{1}{\theta^2 \cdot g \cdot [1 - \beta \cdot \alpha - (1-\beta) \cdot \gamma]} > 0 \tag{17}$$

Higher preparation costs require higher detection probability to sustain the preparation equilibrium. This result aligns with intuition: when studying is more costly, students need stronger deterrents to avoid free-riding.

The effect of $\beta$ (portion completed before swapping) depends on the relative difficulty parameters:

$$\frac{\partial \bar{p}}{\partial \beta} = \frac{-\theta \cdot g \cdot (\gamma - \alpha) \cdot [c/\theta - \theta \cdot g \cdot (1-\alpha)]}{[\theta \cdot g \cdot (1 - \beta \cdot \alpha - (1-\beta) \cdot \gamma)]^2} \tag{18}$$

The sign depends on whether $\gamma > \alpha$. If completing another's script is harder than answering without preparation ($\gamma < \alpha$), increasing $\beta$ (more pre-swap completion) raises the threshold, making the deterrent less effective. Conversely, if $\gamma > \alpha$, the mechanism becomes more effective. This finding suggests an important implementation consideration: the timing of detection affects deterrent strength. Early detection (low $\beta$) may be more effective when students find it particularly difficult to complete others' work ($\gamma$ is low).

## 5.3 Heterogeneous Abilities

Consider students with different abilities $\theta_A > \theta_B$. The equilibrium structure becomes more complex. With sufficient ability heterogeneity, multiple equilibria can exist:

- If $\theta_A/\theta_B > \bar{\rho}$, a separating equilibrium exists where only the high-ability student prepares
- If $\theta_A/\theta_B < \rho$, mutual preparation remains the unique equilibrium
- For intermediate ratios, both pooling and separating equilibria may coexist

where $\bar{\rho}$ and $\rho$ are threshold ratios determined by the model parameters. This multiplicity of equilibria when abilities differ significantly suggests that script swapping may be most effective in relatively homogeneous classrooms—an empirically testable prediction.

## 5.4 Welfare Analysis

To evaluate the welfare implications of the script-swapping mechanism, we compare total surplus under different equilibria and punishment regimes. Define social welfare as:

$$W = \sum_{i \in \{A,B\}} U_i + \lambda \cdot I \tag{19}$$

where $I$ represents the institution's objective (e.g., maintaining academic integrity) and $\lambda$ weights its importance.

**Proposition 4** (Welfare Superiority of Script Swapping). *The script-swapping mechanism achieves higher social welfare than traditional penalties when:*

$$p^{swap} \cdot [W(P, P) - W(NP, NP)] > p^{trad} \cdot [W(P, P) - W^{trad}(NP, NP)] \tag{20}$$

*where superscripts denote the punishment regime.*

The key insight is that script swapping not only deters cheating but also changes the equilibrium preparation levels. Traditional penalties that simply reduce grades may not shift equilibrium behavior if the penalty is not sufficiently severe—a finding consistent with observations that penalties can sometimes be counterproductive if perceived as prices rather than punishments.

# 6 Discussion

Our analysis reveals several fundamental insights about how the script-swapping mechanism operates as a deterrent to academic dishonesty. The central finding is that by creating strategic interdependence between students, the mechanism transforms the incentive structure in ways that differ qualitatively from traditional punishment approaches. This section interprets our results in the context of broader theoretical frameworks and existing empirical evidence.

## 6.1 The Power of Strategic Interdependence

The most striking feature of the script-swapping mechanism is how it creates mutual vulnerability between potential cheaters. Unlike traditional penalties that impose costs only on detected cheaters, script swapping makes each student's outcome dependent on the other's preparation level. This interdependence operates through two channels. First, an unprepared student who copies from a prepared student faces the prospect of completing an exam they are ill-equipped to handle if detected. Second, a prepared student whose work is copied faces the costly prospect of completing another student's largely empty script. This bilateral risk fundamentally alters the cost-benefit calculation that underlies the decision to cheat.

Our formal analysis demonstrates that this interdependence reduces the detection probability threshold required for deterrence compared to traditional penalties. The intuition is that students must now consider not just the probability of being caught, but also the state of the other student's preparation. This additional layer of uncertainty, combined with the asymmetric costs of swapping depending on relative preparation levels, creates a coordination problem that favors mutual preparation as the equilibrium outcome.

## 6.2 Multiple Equilibria and Coordination Challenges

A key finding from our equilibrium analysis is that multiple equilibria can exist under certain parameter configurations, particularly when students have heterogeneous abilities or when preparation costs vary significantly. The existence of multiple equilibria has important implications for understanding observed patterns of academic dishonesty. In some institutional contexts, high-cheating equilibria may persist even with relatively strong detection mechanisms, while in others, honest preparation may be widespread with minimal enforcement.

The mutual non-preparation equilibrium is particularly interesting from a policy perspective. This equilibrium represents a coordination failure where both students would be better off preparing, but unilateral preparation creates vulnerability to exploitation. Traditional deterrent mechanisms may be ineffective at shifting behavior out of this equilibrium because the problem is not primarily about deterring cheating, but about coordinating on the efficient outcome. Script swapping addresses this coordination problem by making it costly for unprepared students to attempt to free-ride on prepared students' work.

## 6.3 Limitations and Boundary Conditions

Our analysis also reveals important limitations of the script-swapping mechanism. The coordinated partial preparation scenario demonstrates that sophisticated forms of collaborative cheating may not be effectively deterred by script swapping. When students can divide the curriculum and each prepare different sections, the swapping mechanism may actually facilitate rather than deter collaborative cheating, as students with complementary preparation can complete each other's scripts relatively successfully.

This finding suggests that script swapping is most effective against opportunistic copying between independently preparing students, rather than against premeditated collaborative cheating schemes. Institutions implementing such mechanisms should therefore view them as part of a comprehensive academic integrity strategy rather than a standalone solution. Complementary measures to detect coordinated cheating patterns, such as analyzing similarities in student preparation choices or monitoring communication patterns, may be necessary.

## 6.4 Behavioral Considerations Beyond the Model

While our model assumes rational, forward-looking agents, real students may respond to script swapping through additional behavioral channels not captured in the formal analysis. The psychological impact of being forced to complete another student's work may be particularly salient. This punishment carries an element of humiliation and public exposure that could amplify its deterrent effect beyond what pure cost-benefit calculations would suggest. Students may experience shame not just from being caught cheating, but from the visible demonstration of their lack of preparation when forced to complete an unfamiliar script.

Additionally, fairness considerations may enhance the mechanism's acceptability. Unlike penalties that simply reduce grades, script swapping creates a form of restorative justice where students face consequences directly related to their transgression. The punishment is neither purely retributive nor purely compensatory, but instead forces students to confront the consequences of their lack of preparation in a concrete way. This may make the mechanism more acceptable to both students and faculty than arbitrary grade reductions or academic suspensions.

## 6.5 Detection Probability and Implementation Timing

Our comparative statics analysis reveals that the timing of detection—captured by the parameter $\beta$ representing the portion of the exam completed before swapping—significantly affects the mechanism's deterrent strength. The relationship between timing and effectiveness depends on the relative difficulty of completing someone else's partially finished work versus completing exam questions without preparation. When completing another's work is particularly difficult (low $\gamma$), early detection is more effective because it maximizes the portion of the exam that must be completed under swapping conditions.

This finding has practical implications for implementation. Institutions should invest in monitoring technologies and proctoring strategies that enable early detection of copying attempts. Real-time monitoring systems, strategic seating arrangements that facilitate observation, and multiple proctors who can quickly identify suspicious behavior all become more valuable under script swapping because they allow detection at earlier stages of the examination.

## 6.6 Cultural and Institutional Context

An important consideration for implementing script swapping is the role of cultural and institutional context. The mechanism has been discussed primarily in South Asian educational settings, where cultural attitudes toward cooperation, competition, and punishment may differ from those in Western contexts. Our model abstracts from cultural factors, but empirical implementation would need to account for how students perceive and respond to this form of punishment in different cultural settings.

Cultural attitudes toward shame, public exposure, and restorative justice may significantly influence the mechanism's effectiveness and acceptability. In cultures where public acknowledgment of failure carries particularly high social costs, the deterrent effect may be amplified. Conversely, in contexts where collaborative learning is highly valued and boundaries between cooperation and cheating are more ambiguous, students may view script swapping as an unfair punishment for behavior they do not perceive as seriously wrong.

# 7 Implications

## 7.1 Policy Implications for Educational Institutions

The findings from our analysis have important implications for educational institutions seeking to combat academic dishonesty through innovative approaches. First and foremost, our results suggest that institutions should pay greater attention to the structure of punishments, not just their severity. The script-swapping mechanism demonstrates that creating strategic interdependence between potential violators can achieve deterrence more efficiently than simply increasing traditional penalties. This insight may extend beyond academic contexts to other enforcement problems where violations involve multiple parties.

For institutions considering implementing script swapping or similar mechanisms, our analysis identifies several critical factors for success. The detection probability threshold identified in Theorem 1 provides guidance on the minimum level of monitoring required for the mechanism to shift equilibrium behavior toward honest preparation. Institutions should assess whether their current detection capabilities meet this threshold and, if not, invest in monitoring technologies and proctoring resources to reach the necessary level. Our comparative statics suggest that such investments in detection may yield higher returns under script swapping than under traditional penalty systems because the enhanced deterrent effect amplifies the value of each incremental increase in detection probability.

The timing of detection emerges as a crucial implementation detail. Early detection maximizes the deterrent effect when students find it particularly difficult to complete each other's work. Institutions should therefore prioritize monitoring strategies that enable quick identification of copying attempts, such as real-time proctoring systems, strategic seating arrangements, and clear protocols for immediate intervention when suspicious behavior is observed. The marginal value of early detection under script swapping may justify investments in monitoring technologies that would not be cost-effective under traditional penalty systems.

## 7.2 Implications for Mechanism Design

From a mechanism design perspective, our analysis contributes to understanding how punishment structures can create self-enforcing incentives for honest behavior. The script-swapping mechanism achieves deterrence not primarily through the severity of the punishment, but through the strategic considerations it induces. This suggests a broader principle: effective deterrent mechanisms should aim to align individual incentives with desired social outcomes through strategic complementarities rather than simply imposing costs on violators.

This principle has potential applications beyond academic integrity. In any setting where violations involve multiple parties and where the success of dishonest behavior depends on others' actions, creating mutual vulnerability through interdependent punishments may enhance deterrence. For example, in corporate settings, making co-conspirators in fraud jointly responsible for remediation rather than imposing independent fines might create stronger incentives for honest behavior. In environmental regulation, making firms in a supply chain collectively responsible for violations at any point in the chain could create peer monitoring and pressure for compliance.

However, our analysis also reveals important limitations that constrain the applicability of interdependent punishment mechanisms. The coordinated partial preparation scenario demonstrates that when parties can credibly commit to complementary roles in dishonest behavior, interdependent punishments may lose their deterrent effect. This suggests that such mechanisms are most effective against opportunistic violations rather than premeditated conspiracies. Institutions implementing interdependent punishments should therefore combine them with measures to detect and prevent coordination among potential violators.

## 7.3 Implications for Student Welfare and Learning

An important consideration often overlooked in discussions of academic integrity enforcement is the impact on student welfare and learning outcomes. Traditional punishments like grade reductions or academic suspension impose costs on students but do not directly promote learning. Script swapping, by contrast, forces students to confront their lack of preparation in a concrete way. While this may be stressful and uncomfortable, it could potentially have educational value by making the consequences of inadequate preparation salient and immediate.

However, this potential benefit must be weighed against fairness concerns. Prepared students who are victimized by cheaters face costs under script swapping that they would not face under traditional penalties. Being forced to complete another student's largely empty script imposes time costs, stress, and potentially affects the prepared student's grade if they cannot complete both scripts adequately. Institutions must carefully consider how to address these fairness concerns, perhaps through mechanisms that provide some grade protection for prepared students who are forced to swap or through accelerated appeals processes that allow wrongly implicated students to quickly resolve their situations.

The distributional effects of script swapping also warrant consideration. If ability heterogeneity is large, our analysis suggests that multiple equilibria may exist, potentially leading to separating equilibria where only high-ability students prepare. This could exacerbate educational inequality if low-ability students rationally choose not to prepare given the costs and expected benefits. Institutions should monitor whether script swapping has differential effects across student populations and consider complementary interventions to support struggling students.

## 7.4 Implications for Honor Codes and Institutional Culture

Our analysis suggests that script swapping may be most effective when embedded within a broader institutional culture that emphasizes academic integrity as a collective responsibility rather than merely an individual obligation. The mechanism creates peer effects by making students' outcomes interdependent, but these peer effects can work in multiple directions. In institutions where cheating is normatively acceptable, script swapping might simply be viewed as an unfortunate risk of getting caught rather than as a serious violation of shared values. In institutions with strong honor code cultures, however, the mechanism may reinforce existing norms by providing a concrete demonstration of how dishonest behavior imposes costs on peers.

This suggests that script swapping should be introduced alongside efforts to cultivate institutional cultures that support academic integrity. Clear communication about the rationale for the mechanism—emphasizing not just the deterrent effect but also the principle that academic dishonesty creates negative externalities for honest students—may enhance its effectiveness. Student involvement in developing and implementing the policy could build buy-in and reinforce the message that academic integrity is a shared responsibility.

## 7.5 Implications for Responsive Regulation and Graduated Sanctions

Our analysis has focused on script swapping as a binary punishment—either it is applied when students are caught copying, or it is not. However, the principles of responsive regulation suggest that graduated sanctions that increase in severity for repeat offenses may be more effective than uniform punishments. Incorporating tapering off mechanisms—where punishment severity is gradually reduced for students who demonstrate improved behavior over time—could enhance the mechanism's effectiveness while also promoting rehabilitation.

For example, institutions could implement a system where first-time offenders face full script swapping, but students who subsequently demonstrate honest behavior through a period of supervised examinations might face reduced penalties for any future violations. This graduated approach acknowledges that some dishonest behavior may reflect poor judgment or desperation rather than fundamental character flaws, and provides pathways for students to rebuild trust. The key is to maintain sufficient deterrence while also creating incentives for reform.

Such graduated systems must be carefully designed to avoid creating perverse incentives. If penalties taper too quickly, students might calculate that the reduced future costs are worth the immediate benefits of cheating. The challenge is to find the right balance between deterrence and rehabilitation—a balance that may vary depending on institutional context, student population characteristics, and the nature of the academic dishonesty problem faced by a particular institution.

# 8 Future Research Directions

The analysis presented in this paper opens several promising avenues for future research that could extend both theoretical understanding and practical applications. This section outlines key directions for future work, organized around empirical validation, theoretical extensions, and policy development.

## 8.1 Empirical Validation and Behavioral Testing

The most immediate priority for future research is empirical validation of the model's predictions. This could take several forms. First, field experiments in educational institutions that implement script swapping would provide direct evidence on the mechanism's effectiveness in real-world settings. Carefully designed natural experiments that compare institutions with and without script swapping policies, or that analyze changes in cheating behavior before and after policy implementation, could identify causal effects using difference-in-differences or regression discontinuity approaches. Such studies should measure not just cheating rates but also preparation effort, student well-being, grade distributions, and long-term learning outcomes to fully assess the mechanism's impact.

Second, laboratory experiments could provide controlled tests of specific theoretical predictions. For example, experiments could manipulate detection probability, preparation costs, and ability heterogeneity to test whether equilibrium behavior shifts in the ways predicted by our model. Laboratory settings allow for precise measurement of beliefs about others' preparation, willingness to engage in copying behavior, and responses to different punishment structures. Experimental designs could directly compare script swapping to traditional penalties holding constant factors like detection probability and punishment severity, isolating the effect of creating strategic interdependence.

Third, survey methods could explore student perceptions of script swapping and how these perceptions relate to behavioral intentions. Do students view script swapping as fair? Do they perceive it as a stronger deterrent than traditional penalties? How do cultural factors influence these perceptions? Survey evidence on these questions would complement behavioral data and provide insights into the psychological mechanisms through which script swapping operates. Combining survey data with behavioral measures would allow researchers to test whether students who perceive script swapping as particularly costly are indeed less likely to engage in cheating behavior.

## 8.2 Theoretical Extensions

Several theoretical extensions could enrich the basic model. First, incorporating repeated interactions and reputation effects would allow analysis of how trigger strategies might sustain cooperation even with relatively low detection probabilities. In settings where students take multiple courses together or where information about past cheating behavior circulates through social networks, reputation concerns could amplify the deterrent effect of script swapping. Models with repeated interactions could also examine how institutions might optimally adjust their monitoring intensity over time as they learn about student behavior and as students update their beliefs about detection probability based on experience.

Second, explicit modeling of network effects would capture how information about preparation strategies and cheating opportunities spreads through student social connec-tions. Following recent research on peer effects in academic dishonesty, future work could examine how network structure—including network density, clustering, and centrality measures—affects the emergence and stability of different equilibria. Network models could also analyze how information cascades might lead to rapid shifts between high-cheating and low-cheating equilibria, and how institutions might strategically intervene in networks to promote honest behavior.

Third, incorporating incomplete information and signaling could provide insights into how students form beliefs about others' preparation levels and how they might strategically signal their own preparation to deter copying attempts. Signaling models could examine whether prepared students have incentives to reveal or conceal their preparation, and how observable pre-exam behaviors (like attending review sessions or asking questions) serve as signals. Such models could also analyze how script swapping affects signaling incentives—for example, whether the mechanism reduces the value of falsely signaling preparation since unprepared students who copy from signaling students face serious consequences if detected.

Fourth, relaxing the binary preparation assumption to allow continuous effort choices would permit more nuanced analysis of partial preparation strategies and how script swapping affects the intensive margin of study effort. Models with continuous effort could examine optimal preparation strategies as a function of beliefs about others' effort, detection probability, and the parameters of the swapping mechanism. Such models might reveal non-monotonic relationships between detection probability and effort that are obscured in binary choice models.

Finally, incorporating explicit behavioral considerations—including loss aversion, shame and guilt, fairness preferences, and present bias—would provide richer predictions about how students actually respond to script swapping. Behavioral models could explore whether the mechanism's effectiveness is amplified or diminished by deviations from expected utility theory, and could generate predictions about heterogeneity in responses based on individual differences in psychological factors. Such models might explain why script swapping appears to be particularly effective in some contexts but not others.

## 8.3 Policy Design and Implementation

From a policy perspective, several important questions merit investigation. First, research on optimal mechanism design could characterize the detection technologies, punishment functions, and information disclosure policies that maximize social welfare while minimizing enforcement costs. Mechanism design approaches could formally derive optimal policies as functions of institutional characteristics like class size, student ability distributions, and the cost of monitoring technologies. Such research could provide concrete guidance to institutions on how to tailor script swapping to their specific contexts.

Second, research on implementation challenges could examine the practical difficulties of deploying script swapping in real educational settings. Qualitative studies of institutions that have experimented with similar mechanisms could document the barriers to effective implementation—including faculty resistance, student complaints, legal challenges, and administrative burdens. Understanding these practical challenges is essential for translating theoretical insights into workable policies. Research could also examine how to design communication strategies that frame script swapping in ways that maximize acceptance among stakeholders.

Third, comparative institutional analysis could evaluate the mechanism's relative ef-fectiveness across different educational contexts, including high-stakes standardized testing, regular course examinations, take-home assignments, and online assessments. The optimal design of deterrent mechanisms likely varies across these contexts depending on factors like the ease of detection, the value of the assessment to students, and the feasibility of implementing swapping. Research comparing effectiveness across contexts would help institutions decide whether and how to adopt script swapping in their specific settings.

Fourth, research on dynamic policy adaptation could model how institutions should adjust their academic integrity policies over time as they learn about student behavior and as students adapt to enforcement mechanisms. Dynamic models could examine optimal experimentation strategies where institutions try different policies and update based on observed outcomes. Such models could also address how to handle strategic student responses to policies, including the possibility that students may find new ways to cheat that circumvent script swapping.

Finally, research on graduated sanctions and tapering mechanisms could explore how to optimally design systems where punishment severity varies with offense history and post-offense behavior. Following the responsive regulation framework, research could examine how to balance deterrence and rehabilitation objectives, how quickly penalties should taper for students showing improvement, and how to avoid creating perverse incentives where students game the system by cycling between cheating and reform. Optimal tapering schedules likely depend on institutional characteristics and the specific academic integrity challenges faced.

## 8.4 Extensions to Other Contexts

The principles underlying script swapping may extend beyond academic settings to other enforcement contexts where violations involve multiple parties and where creating strategic interdependence could enhance deterrence. Future research could explore applications to corporate fraud, environmental violations, tax evasion, and other domains where co-offenders currently face largely independent punishments. Such research could examine whether interdependent punishment mechanisms achieve better outcomes than traditional approaches in these contexts, and could identify the boundary conditions under which such mechanisms are effective.

Research in these domains could also provide insights that feed back into understanding script swapping in academic contexts. For example, studies of how corporate employees respond to joint liability for fraud might reveal behavioral patterns that also apply to students facing script swapping. Cross-domain comparisons could identify general principles about when and how interdependent punishments enhance deterrence, contributing to a broader theory of optimal enforcement mechanism design.

# 9 Conclusion

This paper has presented the first formal game-theoretic analysis of an innovative deterrent mechanism for academic dishonesty: forced script swapping between students caught copying. By creating strategic interdependence between potential cheaters, the mechanism fundamentally transforms the incentive structure surrounding decisions to prepare for examinations and to engage in copying behavior. Our analysis demonstrates that script swapping can achieve deterrence more efficiently than traditional penalties by making students' outcomes mutually dependent on preparation levels, thereby converting what would otherwise be a dominant strategy of copying into a risky proposition that depends critically on beliefs about peer behavior and detection probabilities.

The central theoretical contribution is showing that punishment structure, not just punishment severity, can be a powerful policy instrument for deterring violations. When students caught copying are forced to exchange scripts, they face consequences that depend on both their own lack of preparation and the preparation level of the student they copied from. This bilateral risk creates mutual vulnerability that discourages free-riding behavior. Our equilibrium analysis reveals that mutual preparation emerges as the unique Nash equilibrium when detection probability exceeds a threshold that is notably lower than what traditional penalties would require for equivalent deterrence.

The analysis also reveals important limitations. The mechanism may not effectively deter sophisticated forms of collaborative cheating where students coordinate partial preparation strategies. Additionally, when student abilities are highly heterogeneous, multiple equilibria may exist, potentially including separating equilibria where only high-ability students prepare. These findings suggest that script swapping should be part of a comprehensive academic integrity strategy rather than a standalone solution, and may be most effective in relatively homogeneous educational settings.

From a practical standpoint, institutions considering script swapping should carefully attend to detection probability, implementation timing, and institutional culture. The mechanism requires sufficient monitoring to meet the minimum detection threshold, benefits from early detection when students find completing others' work particularly difficult, and works best when embedded within a culture that emphasizes academic integrity as a collective responsibility. Fairness concerns about costs imposed on prepared students must also be addressed through appropriate safeguards and appeals processes.

Looking forward, the framework developed here provides a foundation for empirical research using field experiments, laboratory studies, and survey methods to validate the model's predictions and explore behavioral mechanisms not captured in the formal analysis. Theoretical extensions incorporating repeated interactions, network effects, incomplete information, and behavioral factors could enrich our understanding of how script swapping operates in complex real-world settings. Policy research on optimal mechanism design, implementation challenges, and applications to other enforcement contexts could translate these insights into practical guidance for institutions.

Ultimately, the script-swapping mechanism exemplifies how innovative thinking about enforcement can yield approaches that align individual incentives with collective goals in novel ways. While not a panacea for academic dishonesty, the mechanism demonstrates the potential for punishment structures that create strategic complementarities between potential violators. As educational institutions worldwide grapple with maintaining academic integrity in an era of evolving technologies and student behaviors, such innovations offer promising directions for achieving deterrence while promoting fairness, learning, and institutional cultures that value honest academic work. By bridging game theory, behavioral economics, and educational policy, this research contributes to a more sophisticated understanding of how institutions can effectively promote academic integrity through well-designed enforcement mechanisms.